\documentstyle[preprint,aps,psfig]{revtex}
\tighten
\draft
\begin{document}
\preprint{\today}
\title{
Test of mode coupling theory for a supercooled liquid of diatomic
molecules.I. Translational degrees of freedom}
\author{Stefan K\"ammerer, Walter Kob and Rolf 
Schilling}
\address{Institut f\"ur Physik, Johannes Gutenberg-Universit\"at,
Staudinger Weg 7, D-55099 Mainz, Germany}
\maketitle

\begin{abstract}
A molecular dynamics simulation is performed for a supercooled liquid
of rigid diatomic molecules.  The time-dependent self and collective
density correlators of the molecular centers of mass are determined and
compared with the predictions of the ideal mode coupling theory (MCT)
for simple liquids.  This is done in real as well as in momentum
space.  One of the main results is the existence of a unique transition
temperature $T_c$, where the dynamics crosses over from an ergodic to a
quasi-nonergodic behavior. The value for $T_c$ agrees with that found
earlier for the orientational dynamics within the error bars. In the
$\beta$-regime of MCT the factorization of space- and time dependence
is satisfactorily fulfilled for both types of correlations. The first
scaling law of ideal MCT holds in the von Schweidler regime, only,
since the validity of the critical law can not be confirmed, due to a
strong interference with the microscopic dynamics. In this first
scaling regime a consistent description within ideal MCT emerges only,
if the next order correction to the asymptotic law is taken into
account. This correction is almost negligible for $q=q_{max}$ , the
position of the main peak in the static structure factor $S(q)$, but
becomes important for $q=q_{min}$, the position of its first minimum.
The second scaling law, i.e. the time-temperature superposition
principle, holds reasonably well for the self and collective density 
correlators and different values for $q$.  The $\alpha$-relaxation 
times $\tau_q^{(s)}$ and $\tau_q$ follow a power law in $(T-T_c)$ 
over 2 -- 3 decades. The corresponding exponent $\gamma$ is 
weakly $q$-dependent and is around $2.55$. This value
is in agreement with the one predicted by MCT from the value of the von
Schweidler exponent but at variance with the corresponding exponent
$\gamma \approx 1.6$ obtained for the orientational correlators
$C_1^{(s)}(t)$ and $C_1(t)$, studied earlier.

\end{abstract}

\narrowtext


\section{Introduction}
\label{sec1}

Although several interpretations of the glass transition exist (see e.g. 
Ref. \cite{ref1}), the only microscopic approach leading to a variety of
predictions is based upon the mode coupling theory (MCT). This theory,
which was proposed first by Bengtzelius, G\"otze and Sj\"olander \cite{ref2} and
Leutheusser \cite{ref3}, provides an equation of motion for the normalized
density correlator 
\begin{equation}
   F(q,t) = \frac{\left< \delta\rho_q^\ast(t)  \delta\rho_q  \right>}
			{\left< \delta\rho_q^\ast  \delta\rho_q \right> } 
\end{equation}
of a simple liquid. Note that $F$ depends on $q=\left|\vec{q}\right|$
only, due to the isotropy of the system. In its idealized version, MCT
predicts the existence of a dynamical transition at a critical
temperature $T_c$ (or a critical density $n_c$) from an ergodic to a
nonergodic phase, corresponding to a liquid and a glass, respectively.
The nonergodicity parameter (NEP)
\begin{equation}
f(q) = \lim_{t\rightarrow \infty} F(q,t)
\end{equation}
serves as an order parameter for that transition. $f(q)$ may change
either continuously (type-A transition) or discontinuously (type-B
transition) at $T_c$. For the structural glass transition only the
latter is relevant.

The test of the predictions of the ideal MCT has challenged both,
experimental work and computer simulations. Most of the experiments
have determined the intermediate scattering function $F(q,t)$ or its
time-Fourier transform, the dynamical structure factor (coherent and
incoherent part).  This yields information on the dynamics of the
translational degrees of freedom (TDOF), only. A great part of these
investigations was reviewed by G\"otze and Sj\"ogren \cite{ref4} and in
Refs.~\cite{mct_review}.

Since most of the glass formers are molecular systems, it is also
important to study the role of the orientational degrees of freedom
(ODOF) and their coupling to the TDOF. A convenient way for this is,
e.g., the use of dielectric spectroscopy. Recent results from dielectric
measurements \cite{ref7} are partly consistent with the predictions of
MCT for simple liquids. The interpretation of light scattering
experiments is not so obvious. Since TDOF and ODOF may both contribute
to light scattering, it is not so easy to separate the orientational
part. (see e.g. Ref. \cite{ref8}).  Concerning computer simulations,
only a few molecular liquids have been studied so far
\cite{ref9,ref10,ref11,ref12,ref13,ref14,ref15,ref16}.  In a recent
molecular dynamics [MD]-simulation \cite{ref17} the present authors
investigated a liquid of diatomic, rigid molecules with
Lennard-Jones interactions.  Apart from linear molecules with head-tail
symmetry, this is the simplest choice for a molecular system. Because
dielectric and light scattering measurements yield essentially only
information for $q\approx 0$, we restricted ourselves to
orientational correlators with $q=0$ and to the translational and
rotational diffusion constant $D$ and $D_r$, respectively \cite{ref17}.
The main result we have found is the existence of a power law
dependence of all the orientational relaxation times $\tau_l^{(s)}
(l=1-6)$, $\tau_l$ ($l=1$) and of $D$ on temperature with a single critical
temperature $T_c = 0.475$, in agreement with ideal MCT. Slightly 
above $T_c$ the numerical results for $\tau_l^{(s)}$, $\tau_l$ and $D$
deviate from a power law, due to ergodicity restoring processes. These
processes, which are not considered by the ideal MCT, can be accounted
for by its extended version \cite{ref18}. The observation that the
temperature dependence of $D_r$ does not fit into this scheme, but
bifurcates from $D$ at a temperature significantly above $T_c$ and then
obeys an Arrhenius law, rather than a power law, has been a further
important result.

The main purpose of the present and the following paper \cite{ref18a}
is the extension of our recent results \cite{ref17} to finite $q$  and
to perform a thorough test of the predictions of the ideal MCT. The
present paper is restricted to the TDOF only, whereas the following
paper
\cite{ref18a} considers orientational correlators, which explicitly
contain the coupling between ODOF and TDOF. Although our simulation
investigates a molecular system, we make a comparison with the
predictions of MCT for {\it simple} liquids. The ideal MCT for simple
liquids has been recently extended to a dumbbell molecule in a simple
liquid \cite{ref19} and to molecular liquids of linear
\cite{ref20,ref21} and arbitrary shaped molecules \cite{ref22}. So far,
the investigations of these molecular MCT-equations were restricted to
the calculation of the NEP \cite{ref19,ref21,ref22}.
 For linear molecules without head-tail symmetry, it follows from these
equations \cite{ref21} that the TDOF and ODOF freeze at a single
critical temperature $T_c$, consistent with our recent results for
$q=0$ \cite{ref17}. The investigation of the time dependent molecular
MCT-equations, which would allow comparison with our MD-results, will
be done in future.

The outline of this paper is as follows. The next section will review
the predictions of the ideal MCT (for simple liquids) which will be
tested. In section~\ref{sec3} the model as well as some details of the
computer simulation are discussed.  In section~\ref{sec4} we present
our MD-results and the final section contains a discussion of these
results and our main conclusions.

\section{mode-coupling theory}
\label{sec2}

In this section we will give a short summary of those predictions of
the ideal MCT (for simple liquids), which will be compared with our
MD-results. For details the reader may consult the review papers 
\cite{ref4,mct_review}. For completeness we also would like
to mention the MCT-approach by use of nonlinear fluctuating
hydrodynamics \cite{ref25,ref26}.

Ideal MCT predicts the existence of a dynamical transition at $T_c$ from an ergodic to a
nonergodic phase which is an ideal glass transition. For temperatures
close to $T_c$, MCT predicts the existence of two scaling laws for $F(q,t)$
with time scales $t_{\sigma}$ and $\tau (\gg t_\sigma)$ , where $\tau$ is the 
$\alpha$-relaxation time. In the first scaling regime, i.e. for
$t_0 \ll t \ll \tau$, the density correlator takes the form:
\begin{equation}
F(q,t) = f^c(q) + h(q) G(t)
\end{equation}
with $f^c(q)$ the NEP at $T_c$ and $h(q)$ the critical amplitude.  $t_0$ is a
microscopic time of the order of $10^{-13}$ sec. The so-called
$\beta$-correlator (or critical correlator) $G(t)$ obeys the first scaling law:
\begin{equation}
G(t) = c_\sigma g_\pm(t/t_\sigma)
\end{equation}
with the correlation scale 
$c_\sigma = \left| \sigma \right|^{1/2}$ and the separation parameter
$\sigma=\sigma_0(T-T_c)$, where $\sigma_0>0$.  The $\sigma$-independent
master functions $g_\pm(\sigma \stackrel{>}{<} 0)$ are solutions of a
certain scaling equation. The corresponding $\beta$-relaxation time
scale $t_\sigma$ is given by:

\begin{equation}
t_\sigma \propto \left| T-T_c  \right|^{-\frac{1}{2a}} \quad , \quad T \stackrel{>}{<}  T_c
\end{equation}
For $G(t)$ one obtains the following asymptotic power laws:
\begin{equation}
G(t) \propto \left\{ \begin{array}{r@{\quad,\quad}l} 
	t^{-a} & t_0 \ll t \ll t_\sigma \\
	-t^{b} & t_\sigma \ll t \ll \tau \\
			\end{array} \right.
\end{equation}
The critical law, upper equation of (6), holds above and below $T_c$, 
whereas the von Schweidler law, lower equation of (6), is valid only
above the transition point. Below $T_c$ , $G(t) $ decays to a constant
for $t \gg t_\sigma$. Both exponents $a$ and $b$ are related to the
exponent parameter $\lambda (0<\lambda<1)$  by:
\begin{equation}
\frac{\Gamma(1-a)^2}{\Gamma(1-2a)} = \lambda =
\frac{\Gamma(1+b)^2}{\Gamma(1+2b)} \quad ,
\end{equation}
with $\Gamma$ the Gamma function. From (7) one gets that $0 < a < \frac{1}{2}$ and
$0 < b \leq 1$. $\lambda$ is determined by the static correlators at $T_c$. The result
(3) which states that in the first scaling regime the $q$- and
$t$-dependence factorizes is one of the most important predictions of the
ideal MCT. Due to this factorization, Eq. (3) can easily be transformed
to real space:
\begin{equation}
\phi(r,t) = f^c(r) + H(r) G(t)
\end{equation}
where $\phi(r,t)$, $f^c(r)$ and $H(r)$ are the Fourier transform of $F(q,t)$, 
$f^c(q)$ and $h(q)$, respectively.

It is important to realize that the result (6) only holds
asymptotically for $T \rightarrow T_c$. The next order correction to both of these
asymptotic laws was recently derived and calculated for a system of
hard spheres by Franosch {\it et al.} \cite{ref27}.  Corrections to the von
Schweidler law \cite{ref28}
\begin{equation}
F(q,t) = f^c(q) - h(q) (t/\tau)^b + h^{(2)}(q) (t/\tau)^{2b} - \ldots
\end{equation}
have already been studied for hard spheres by Fuchs {\it et al.} \cite{ref29}. There it
was demonstrated that these corrections may be important. The
expansion (9) is valid for $t_\sigma \ll t \ll \tau$, where the 
$\alpha$-relaxation time scale $\tau$ is given by
\begin{equation}
\tau(T) \propto (T-T_c)^{-\gamma} \quad , \quad T \ge T_c
\end{equation}
with
\begin{equation}
\gamma = \frac{1}{2a} + \frac{1}{2b} 
\end{equation}
The translational diffusion constant $D$ is predicted to scale like 
$\tau^{-1}$. Therefore it is:
\begin{equation}
D(T) \propto (T-T_c)^{\gamma} \quad , \quad T \ge T_c \qquad .
\label{mct35}
\end{equation}
Corrections to the critical law lead to \cite{ref27}:
\begin{equation}
F(q,t) = f^c(q) + h(q) (t/t_\sigma)^{-a} + \bar{h}^{(2)}(q) (t/t_\sigma)^{-2a} + \ldots
\end{equation}
In the second scaling regime, i.e. for $t$ of the order of $\tau$, a master
function $\tilde{F}_q(\tilde{t})$ exists such that
\begin{equation}
F(q,t,T) = \tilde{F}_q(t/\tau(T))
\end{equation}
In glass science the result (14) is called time-temperature
superposition principle.  The expansion of the r.h.s. of (14) with
respect to $t/\tau$ yields (9). Equation (14) represents the second scaling law.

Finally we mention that all these MCT-results hold for the self part of
the density correlator as well.

\section{Model and details of the simulation}
\label{sec3}

The model we investigate is a one-component system of rigid diatomic
molecules. Each molecule is composed of two different Lennard-Jones
particles, in the following denoted by $A$ and $B$, which are separated
by a distance $d = 0.5$ and  each of which has the same mass $m$. The
interaction between two molecules is given by the sum of the
interaction between the four particles which is given by the
Lennard-Jones potential $V_{\alpha\beta}(r)=4\epsilon_{\alpha\beta}
\{(\sigma_{\alpha\beta}/r)^{12}-(\sigma_{\alpha\beta}/r)^{6}\}$ where
$\alpha,\beta \in \{A,B\}$. The Lennard-Jones parameters are given by:
$\sigma_{AA}=\sigma_{AB}=1.0$, $\sigma_{BB}=0.95$,
$\epsilon_{AA}=\epsilon_{AB}=1.0$ and $\epsilon_{BB}=0.8$. In the
following we will use reduced units and use $\sigma_{AA}$ as the unit
of length, $\epsilon_{AA}$ as the unit of energy (setting $k_B=1$) and
$(\sigma_{AA}^2m/48\epsilon_{AA})^{1/2}$ as the unit of time.  If the
atoms are argon-like this time unit corresponds to approximately 0.3 ps.

In order to make the simulation more realistic we did it at constant
external pressure $p_{ext}$=1.0.  The length of the equilibration runs
always exceeded the typical relaxation time of the system at the
temperature considered, which allows us to conclude that in the
subsequent production runs we were investigating the {\it equilibrium}
dynamics of the system.  The temperatures we investigated are $T=5.0$,
3.0, 2.0, 1.4, 1.1, 0.85, 0.70, 0.632, 0.588, 0.549, 0.520, 0.500,
0.489 and 0.477. The total number of molecules was 500 and, in order to
improve the statistics of the results, we averaged at each temperature
over at least eight independent runs.  For more details see
\cite{ref17}.

\section{Results}
\label{sec4}

For a clearer presentation of our results, this section is divided into
two subsections, where the first is devoted to the static and the
second to the dynamical properties.  The latter is divided into two
parts again, which are real space and $q$-space representation.

\subsection{Static properties}

First of all we remark that thermodynamic quantities like,
e.g. the average density and the enthalpy do not exhibit any
signature of a singular behavior for the full temperature range 
$0.477 \le T \le 5.0$, we have investigated. From this we conclude
that the observed slowing down is probably not related to an approach
of the system to a critical point of a second order phase transition.

The structural properties are one of the most interesting static
features of a supercooled liquid. Figure 1 shows the static structure
factor $S(q)$ of the center of mass positions for different temperatures
as a function of $q$. Its $q$-dependence has the typical behavior 
expected for a liquid, with a main peak at $q_{max}=6.5$ and a first
minimum at $q_{min}=8.15$ for the lowest temperature. With increasing
temperature the peak positions shift to smaller $q$ values, due to an
increase of the average distance between the molecular centers, and the
peaks become broader.

A comparison of $S(q)$ for $T=0.477$ with the partial structure
factors $S_{\alpha \beta}(q)$, where $\alpha,\beta$ refers to atom $A$
and atom $B$, is given in Fig.~2.  Whereas the main peak in $S(q)$ and
$S_{\alpha \beta}(q)$ is essentially at the same position, the
remaining $q$-dependence is quite different for the various
correlators. For the pre-peak of $S_{AA}(q)$ and $S_{BB}(q)$ at $q
\approx 3.1$, which could indicate a ``medium'' range order, we have
not found a conclusive interpretation. Concerning the microscopic
structural properties it is more convenient to consider the atomic
radial pair distribution functions $g_{\alpha \beta}(r)$ and the
corresponding function $g_{cm}(r)$ for the center of mass positions,
which are shown in Figs.~3 and 4, respectively. There is no significant
difference between the $g_{\alpha \beta}(r)$ probably due to the fact
that the choice of our model parameters \cite{ref17} does not introduce
a strong asymmetry between particles of type $A$ and particles of type
$B$.  Besides a first and a splitted second nearest neighbor peak at
about $r_1 \approx 1.05$ and $r_2 \approx 1.9$, $r_2' \approx 2.1$,
respectively (Fig.~3), an intermediate peak occurs at $r_1' \approx
1.5$. The peak at $r_2'$ is also found for $g_{cm}$, but its main peak
is at $r_1'' \approx 1.25$ (cf. Fig. 4). The peaks in $g_{\alpha
\beta}$ at $r_1'$ and $r_2$ are missing for $g_{cm}$.  The splitted
{\it first} nearest neighbor peak of $g_{cm}$ can be attributed to
cross- and T-configurations, as also found for other molecular liquids
(see e.g. \cite{ref30}). We do not dwell  more on these results, since
our main concern are the dynamical features in the strongly supercooled
regime. The main reason we have presented $S(q)$ is to compare its
$q$-dependence with the static correlation functions of the
orientational correlators presented in part II. We already anticipate
that this $q$-dependence can be rather different for the various
correlators.

\subsection{Dynamical properties: real space}

To start we discuss the self motion of a molecule. One of the important
transport coefficient is the translational diffusion constant $D$ which
is obtained from the mean squared displacement by:
\begin{equation}
D = \lim_{t \rightarrow \infty} \frac{1}{6tN} \sum\limits_{n=1}^{N} \left<
			( \vec{x}_n(t) - \vec{x}_n(0) )^2	\right> \quad ,
\end{equation}
where $N$ is the number of molecules.  The temperature-dependence of
$D$ is presented in Fig.~5 on a double-logarithmic scale. The diffusion
constant follows a power law, as predicted by ideal MCT [cf. Eq. (12)],
over a surprisingly large range of four decades in $D$. As critical
temperature we obtain $T_c = 0.475$ and for the corresponding exponent
$\gamma_D = 2.20$.  With use of Eqs.~(7) and (11), this yields the von
Schweidler exponent $b_D = 0.69$, the critical exponent $a_D
= 0.34$ and the exponent parameter $\lambda_D = 0.67$. At
the two lowest temperatures the deviations of the numerical values for
$D$ from the power law, which has been mentioned in the Introduction,
can clearly be seen. The inset of Fig.~5 depicts the time dependence of
the  mean squared displacement for all temperatures, from which $D(T)$
was deduced.

More information on the self motion is contained in the self part of the van
Hove correlator:
\begin{equation}
G_s(r,t) = \left< \frac{1}{N} \sum_{n=1}^{N} \delta(r - \left| \vec{x}_n(t) - 
			\vec{x}_n(0) \right| ) \right>
\end{equation}
The $r$-dependence of $G_s(r,t)$ is shown in Fig.~6 for times which are equidistant 
on a logarithmic time axis
and for the lowest temperature. With time, the $\delta$-peak at $t=0$
broadens and for $t \rightarrow \infty$ and $r \rightarrow \infty$, 
i.e. in the hydrodynamic limit, it approaches a Gaussian distribution:
\begin{equation}
G_s(r,t) \stackrel{t,r \rightarrow \infty}{\longrightarrow} 
\frac{1}{(4 \pi D t)^{3/2}} \exp(-\frac{r^2}{4Dt}) 
\end{equation}
In the time span $5.0 \le t \le 10^3$, where the mean squared
displacement for the lowest temperature exhibits a plateau (cf. the
inset of Fig. 5), the $r$-dependence of $G_s(r,t)$ varies only weakly
with time, which indicates the validity of the factorization [Eq.~(8)].
To check this, we have plotted in Fig.~7 $(G_s(r,t)-G_s(r,t')) /
(G_s(r',t)-G_s(r',t'))$ as a function of $r$ for fixed $r'$ and $t'$
and various $t$-values. If Eq.~(8) is valid, this ratio must be
independent of $t$ and $t'$. For the times $t$ and $t'$ we have chosen
in Fig.~7, this is indeed reasonably fulfilled.  That this is not true
for the full time span, can be inferred from the inset of Fig.~7.

Another interesting conclusion can be drawn from the $r$-dependence of
$G_s$ for the largest time. As can be seen from Fig.~6, at no time
there is any indication for the presence of a secondary peak at
$r\approx1$, the nearest neighbor distance. From the absence of such a
peak it is usually concluded, see e.g. Ref.~\cite{ref31}, that no
hopping processes are present. But for the present system we have
strong evidence, see Fig.~5, that at the lowest temperature hopping
processes are indeed present. Thus we conclude that hopping processes
do not necessarily lead to a secondary peak in $G_s$.

The $r$-dependence of $G_s$ is not Gaussian for a large time regime. The
deviation from a Gaussian can be quantified by the non-Gaussian
parameters $\alpha_n(t) , n=2,3,\ldots$ \cite{ref32}. Fig. 8 shows
\begin{equation}
\alpha_2(t) =  \frac{3 \left< \vec{r}\,^{4}(t) \right>}
		{5 \left< \vec{r}\,^2(t) \right>^2} - 1 \quad . 
\end{equation}
In case of a Gaussian process $\alpha_2(t)$ vanishes. For $t
\rightarrow 0$ and $t \rightarrow \infty$ $\alpha_2(t)$ goes to zero.
But in between there exists a time regime for all temperatures where
$\alpha_2$ is substantially different from zero. The increasing part of
$\alpha_2$ seems to fall onto a master curve, as already found earlier
\cite{ref14,ref31}.  The time $t_{max}(T)$, where $\alpha_2(t)$ obtains
its maximum, obeys a power law $(T-T_c)^{-\gamma_\alpha}$ (see inset of
Fig.~8) with $\gamma_\alpha \approx 2.1$ which coincides with
$\gamma_D$, derived from $D(T)$, in agreement with the results of
Sciortino {\it et al.} \cite{ref14}. This suggests that $\alpha_2(t)$
for $t \approx t_{max}$ is determined by the $\alpha$-relaxation as it
follows from ideal MCT.  Exactly the same conclusions also hold for
$\alpha_3(t)$.

The collective dynamics can be studied by use of the distinct part of
the van Hove correlator
\begin{equation}
G_d(r,t) = \left< \frac{1}{N-1} \sum_{n \neq m} \delta(r - \left| 
\vec{x}_n(t) - \vec{x}_m(0) \right| ) \right>
\end{equation}
which is shown in Fig.~9 for various times and again for the lowest
temperature. Its $r$- and $t$-dependence is quite similar to that found
for a binary liquid \cite{ref31}. The test of the factorization
[Eq.~(8)] can be done as for the self part. The corresponding ratio is
presented in Fig.~10 for the same time regime as in Fig.~7.  Again a
reasonable collapse onto one master curve occurs, in agreement with
ideal MCT.

\subsection{Dynamical properties: $q$-space}

Although the visualization of local structural properties favors the
use of a real space representation of the correlators, most
experimental results are obtained in $q$-space. In addition the
translational invariance of the interactions also suggests a
theoretical description by use of $q$-dependent correlators. For instance
the memory part of the MCT-equation for the density correlator $F(q,t)$ is
``diagonal'' in $q$-space but involves a convolution in real space.

Let us start with the self part $F_s(q,t)$ of the collective density
correlator $F(q,t)$ [cf. Eq.~(1)], which is presented in Fig.~11 for
$q=q_{max}=6.5$ , the position of the main peak in $S(q)$ (cf.
Fig.~1).  We remind the reader that $q$ is measured in units of
$\sigma_{AA}$, the position of the Lennard-Jones minimum for the
$A$-particles. The small bump in $F_s(q,t)$ at $t_s \cong 12$, which is
also present in $F(q,t)$ (cf. Fig.~14 and 15), is due to a sound wave
and $t_s$ is the time for the sound wave to traverse the box size of
our  sample. This effect was also observed by Lewis {\it et al.}
\cite{ref11} and has recently been shown to be much more pronounced in
strong glass formers~\cite{horbach96}.

For the highest temperature, $T=5.0$, the decay of $F_s(q,t)$ is
essentially exponential.  With decreasing temperature the relaxation
crosses over into a two-step process, as predicted by MCT. For the
lowest temperature, $T = 0.477$, a ``quasi-plateau'' can be seen. Its
height is a measure of the critical NEP $f_s^c(q)$. The reader should
note that $F_s(q_{max},t)$ decays to zero for large times, indicating
that the length of the runs is large enough to equilibrate the system.
In order to test MCT, we have determined the $\alpha$-relaxation times
$\tau_q^{(s)}(T)$ from the condition $F_s(q,\tau_q^{(s)})=1/e$. Its
temperature dependence is shown in Fig.~12 (squares) for
$q=q_{max}=6.5$ and $q=q_{min}=8.15$. Using $T_c$ = 0.475, as
determined from $D(T)$, both relaxation times follow a power law over
about three decades. The corresponding $\gamma$-values are practically
the same $\gamma^{(s)} = 2.56$, but differ significantly from $\gamma_D
= 2.20$. Such a discrepancy between $\gamma_D$ and $\gamma^{(s)}$ has
been observed already before~\cite{ref31} and indicates that the MCT
prediction, that the two exponents should be equal, is not valid for
these systems.  $\gamma^{(s)} = 2.56$ yields the von Schweidler
exponent $b=0.55$ and the exponent parameter $\lambda = 0.76$.  We also 
note that a power law fit
with $T_c$ as a free parameter yields a value for $T_c$ which coincides
with $T_c = 0.475$ to within $2 \%$.

The relaxation time $\tau_q^{(s)}(T)$ can now be used to rescale time
in order to test the time-temperature superposition principle
[Eq.~(14)], which is done in Fig.~11b. We find that the relaxation
curves fall indeed onto one master curve for the lowest temperatures.
The late stage relaxation can be fitted well by a
Kohlrausch-Williams-Watts-law (KWW), i.e.  $F_s(q,t) =  A \exp \left[ -
(t/\tau^{(s)})^\beta \right]$, dashed curve.  For higher temperatures
(cf. e.g. the curve for $T = 5.0$) no such scaling exists. We also
note that such a scaling was not possible for certain types of {\it
orientational} correlation functions~\cite{ref17}, which shows that
this prediction of MCT is not a trivial one.

Having demonstrated the validity of the second scaling law [Eq.~(14)],
we can test whether the von Schweidler law, including the next order
correction [Eq.~(9)] fits the master curve well in the late
$\beta$-relaxation regime. As can be seen from Fig.~11b this type of
fit works very well (dotted line). In practice, this has been done for
$F_s(q,t)$ at the lowest temperature by using $b=0.55$ (deduced from
$\gamma^{(s)}$).  The wave vector dependence of $f_s^c(q)$,
$\tilde{h}^s(q)$ and $\tilde{h}^{(2)s}(q)$ is shown in Fig.~13. Here a
comment is in order. The fit of the data with the von Schweidler law 
including the next order correction yields $\tau^{-b}h^s(q)$ and
$\tau^{-2b}h^{(2)s}(q)$ [cf. Eq.~(9)]. Since the $q$-independent
$\alpha$-relaxation time scale $\tau(T)$ can only be determined up to a
$T$-independent factor, the same is true for $h^s(q)$ and $h^{(2)s}(q)$. In
Fig.~13 we therefore show $\tilde{h}^s(q) = \tau^{-b} h^s(q)$ and
$\tilde{h}^{(2)s}(q)=\tau^{-2b} h_q^{(2)s}(q)$.
The $q$-variation of these quantities is quite similar  to that for
hard spheres \cite{ref29}. Particularly we also find a zero of
$\tilde{h}^{(2)s}(q)$ at a finite q-value $q_0 \approx  3.5$ with
$\tilde{h}^{(2)s}(q) <0 (>0)$ for $q<q_0 (q>q_0)$.  Taking into account
that $q$ for the hard sphere system is given in units of the diameter
$d^{HS}=1$, we can deduce $q_0^{HS} \approx 12$ from
Ref.~\cite{ref29}.  That this is about three times larger than our
value, is partly due to a larger effective diameter $d_{eff} \approx
1.5$ of our molecules, compared to $d^{HS}=1$. 

We now turn to the collective dynamics as obtained from the density
correlator $F(q,t)$, which is presented, respectively, for $q=q_{max}$
and $q=q_{min}$ in Figs.~14a and 14b. The time-dependence looks similar
to that of $F_s(q,t)$, but the height of the ``quasi-plateau'' is much
lower and the stretching of the relaxation is more pronounced for the
case of $q=q_{min}$.  The $\alpha$-relaxation time $\tau_q(T)$ has 
been
determined from the condition $F(q,\tau_q)= 0.1$, since using the
$e^{-1}$ definition of $\tau$ would lead to an underestimation of the
$\alpha$--relaxation time for $q=q_{min}$, since the plateau is so
low.  (We note that choosing $e^{-1}$ instead of $0.1$, for the case
where both definitions can be used, or even using a KWW-fit to deduce
$\tau_q(T)$, leads to essentially the same temperature-dependence). The
temperature dependence of $\tau_q$ is shown in Fig.~12.  Taking again
$T_c$ = 0.475 as given, $\tau_q(T)$ obeys a power law, with
$\gamma_{q_{max}} = 2.57  \approx \gamma^{(s)}$ and
$\gamma_{q_{min}}=2.47$.  It is remarkable that the range for the power
law for $q=q_{min}$ is almost one decade less than for $q=q_{max}$,
which shows that different correlators reach the asymptotic regime at
different temperatures.

Fig.~15 shows that also for these correlation functions the second
scaling law holds and that it holds better for $q_{max}$ than for
$q_{min}$. The late stage relaxation can again be fitted well by a
KWW-law (dashed line).
Since the curves for the different temperatures fall onto a master
curve, it is sufficient to focus on the curve for the lowest temperature
in order to test whether the first scaling law holds.  The results of
our analysis are shown in Fig.~16 and 17 for $q=q_{max}$ and
$q=q_{min}$, respectively. Let us discuss $q=q_{max}$ first. As can be
seen from Fig.~16, the von Schweidler law, with $b = 0.55 \; \hat{=} \;
\lambda = 0.76$, the same value we used to fit the von Schweidler law
to $F_s(q,t)$, fits the data over about 2.5 decades in time (long
dashed line). Taking into account the next order correction (cf. Eq.~9)
leads to a significant improvement of the fit for $t \geq 4\cdot10^4$
(solid line). We have also fitted the numerical data with the 
$\beta$-correlator with (dotted line) and without (short dashed line)
correction. 
This was done by solving the equation for $g_{-}(t/t_{\sigma})$ (see 
Ref.~\cite{mct_review}) and making use of Eqs. (3) and (4). Here we 
encounter the same problem as we did for the fit with the von Schweidler 
law, because the $\beta$-relaxation time $t_{\sigma}$ can only be 
determined up to a $T$-independent factor. In order to proceed we have chosen 
for $t_{\sigma}$ the position of the inflection point of 
$F(q_{max},t)$, indicated in Fig.~16 by a filled circle. The optimum values 
$\lambda = 0.76$ and $t_{\sigma}=69$ stemming from that fit were used for the 
similar fit of all the other correlators of the present paper and 
of Ref.~\cite{ref18a}, except for the orientational correlators with $l=1$
\cite{ref18a}.
From the figure we recognize that in the late
$\beta$-relaxation regime these fits are identical to the ones of the
von Schweidler law, as it should be. In the early $\beta$-relaxation 
regime, however, the
$\beta$-correlator fits the data much better than the von Schweidler
law, since part of the approach to the plateau is fitted well also.
Nevertheless, the critical law is not really observed, even if one
includes the corrections to it [cf. Eq.~(13)].

For $q=q_{min}$ it is not possible to obtain a good fit with the von
Schweidler law alone, if $b$ is kept fixed at 0.55. If $b$ is used as a
free parameter, a satisfactory fit is obtained but at the cost of a
$q$-dependent von Schweidler exponent $b_q$, in contradiction to MCT.
Since the critical NEP obtained from this fit were rather structureless
and did not qualitatively agree with the NEP obtained from solving the
molecular MCT-equations \cite{ref22}, we decided to keep $b$ = 0.55
(obtained from $\gamma^{(s)}$) fixed and to take the next order
correction into account. With this approach we obtained good agreement
between all NEP from our MD-simulation and those from the MCT-equations
\cite{ref22}. But even if the first correction to the von Schweidler
law (cf. Eq.~9) is taken into account, the fit with Eq.~9 works well
for only about one decade in time. On the other hand, the critical
correlator (with $\lambda=0.76$) including the correction (on the von
Schweidler side) fits the data over two decades. Nevertheless, this
range is significantly smaller than the one found for $q=q_{max}$ which
shows that corrections to the scaling laws might be more important for
$q_{min}$ than for $q_{max}$.

The quantities $f^c(q)$, $\tilde{h}(q)$ and $\tilde{h}^{(2)}(q)$,
depicted in Fig.~18 were determined in a similar way as for the self
part, i.e. by keeping the value of $b$ fixed to  0.55.  Comparing these
quantities with $S(q)$ (Fig. 1) we find that $f^c(q)$ is in phase and
both, $\tilde{h}(q)$ and $\tilde{h}^{(2)}(q)$ in anti-phase with $S(q)$.  We
note that the peak at $q \approx 3$ in $f^c(q)$ does not exist for
$S(q)$. However, whether this peak is real or just a statistical
fluctuation can presently not be decided for sure. Since the static
correlator $S_{11}^0(q)$ has a pronounced maximum at $q\approx 3$ (see
Ref.~\cite{ref18a}), one might be tempted to relate this prepeak of
$f_q^c$ to the translation-rotation coupling.

The variation of $f^c(q)$ and $\tilde{h}(q)$ with $q$ resembles the one
found, e.g., for hard spheres \cite{ref29}, for a binary liquid
\cite{ref31,ref33} and that for water molecules \cite{ref34}.
Interestingly the $q$-dependence of $\tilde{h}^{(2)}(q)$, which is in
phase with that of $\tilde{h}(q)$, qualitatively agrees with that found
by Fuchs {\it et al.} for hard spheres \cite{ref29}. However, in
contrast to the hard sphere system, $\tilde{h}^{(2)}(q)$ does not
change sign for the $q$-regime we have studied, but vanishes at
$q=q_{max}$ at which it has a minimum.  This fact explains why the
(asymptotic) von Schweidler law fits the data rather well for
$q=q_{max}$, where $\tilde{h}^{(2)}(q) \approx 0$, but not for
$q=q_{min}$. To conclude we show in Fig.~19 that also the variation of
the $\alpha$-relaxation time $\tau_q$ with $q$ is in phase with $S(q)$,
which is in close analogy with, e.g., the hard sphere system
\cite{ref29}.

\section{Discussion and conclusions}
\label{sec5}

In this paper we have performed a detailed test of the predictions of
ideal MCT (for simple liquids) for the translational degrees of freedom
(TDOF) for a supercooled liquid of rigid diatomic molecules without
head-tail symmetry. The Lennard-Jones interactions we used for the
molecular system are not so different from those of the binary liquid
\cite{ref31} with $80 \%$ A- and $20 \%$ B-atoms. The main difference
between both systems of course is, that for the present case the $50
\%$ A- and $50 \%$ B-atoms are pairwise connected in order to form
diatomic molecules. Therefore the comparison of the dynamical
behavior of both systems, allows to discuss the influence of the
orientational degrees of freedom (ODOF) on TDOF.

Our test was mainly concerned with (i) the existence of a single
critical temperature $T_c$ and (ii) the validity of the two scaling
laws of ideal MCT. This single critical temperature indicates a strong
coupling between TDOF and ODOF. Although the mathematical structure of
the MCT-equations for molecular liquids \cite{ref19,ref20,ref21,ref22}
differs from the one of simple liquids, we expect that the two scaling
laws hold for molecular liquids, too. This belief is based upon the
fact that the first scaling law is always valid in a so-called type B
transition (i.e. the NEP changes {\it discontinuously} at the
transition), the type of transition that is relevant for structural
glasses~\cite{mct_review}.  Furthermore the existence of a single transition
temperature for linear and arbitrary molecules without additional
symmetry can also be proved for the molecular MCT-equation
\cite{ref35}.

Taking $T_c$ and $\gamma$ as free parameters, the $\alpha$-relaxation
times $\tau_q^{(s)}$ and $\tau_q$ for $q=q_{max}$ and $q=q_{min}$ can
be fitted with a power law.  The resulting $T_c$'s differ from $T_c$ =
0.475 (deduced from $D(T)$) by less than $2 \%$. The same value of
$T_c$ was found for the orientational correlation functions in
Ref.~\cite{ref17} and we therefore conclude that our data is compatible
with the existence of a single transition temperature. The sharp
transition at $T_c$, as predicted by {\it ideal} MCT, is however
smeared out, due to ergodicity restoring processes which can be
accounted for by the extended MCT \cite{ref18}. These processes are
often associated with a hopping of the atoms or molecules, as it has
been demonstrated for a binary liquid \cite{ref36}. An evidence for
such hopping processes is the occurrence of a second peak in $G_s(r,t)$
at $r \approx 1$ for large times~\cite{ref36}. But no such peak is
observed in our results, even beyond the $r$-range shown in Fig.~6. The
absence of hopping processes with respect to the TDOF was also found in
the MD-simulation of CKN \cite{ref9}.  Therefore only orientational
jumps remain, which indeed have been shown to be present \cite{ref17}.
Comparing the different values of $\gamma$, one finds that those
determined from $F_s(q,t)$ and $F(q,t)$ fluctuate around 2.55. The same
is more or less true  for the purely orientational correlators
$C_l^{(s)}(t)$ and $C_l(t)$ [17] for $l>1$, but not for $l=1$ where
$\gamma_1^{(s)} =1.66$ and $\gamma_1 =1.52$ was found. This result
demonstrates that, contrary to the prediction of ideal MCT, $\gamma$ is
not a ``universal'' exponent for our system. Such a conclusion was
already drawn for binary liquids \cite{ref31}. We note however, that
this discrepancy does not seem to exist for the MD-simulation of water
\cite{ref14}.

For all the correlators related to the TDOF, the second scaling law is
reasonably well fulfilled. This is in contrast to our findings for
$C_1^{(s)}(t)$ and $C_1(t)$~\cite{ref17}. Only for relatively 
large values of $l$ do the $C_l^{(s)}(t)$ show the second scaling law.

Regarding the first scaling law we can say that it works well for
$F_s(q,t)$ and $F(q,t)$. For $q=q_{min}$ we have demonstrated that the
next correction to the von Schweidler law becomes important. We stress
that a consistent description of our data in the framework of MCT is
only possible by taking this correction into account. If this is not
done, a $q$-dependent von Schweidler exponent and a rather
structureless $q$-variation of the orientational NEP (which will be
discussed in the following paper~\cite{ref18a}) results. The critical
law is not observed for  any of the correlators, even if the next
correction is taken into account.  This is probably due to a strong
interference of the critical dynamics with the microscopic motion.

In summary, we conclude that, with respect to ideal MCT, the TDOF of
our molecular system behave quite similar as found for a binary liquid
\cite{ref31}. Hence, the qualitative features of the dynamics of TDOF
is not altered by its coupling to ODOF. Although a single transition
temperature for TDOF and ODOF can be specified, part of the
orientational dynamics as measured, e.g. by $C_1^{(s)}(t)$, does not
fit into the framework of ideal MCT \cite{ref17}. Whether this is only due to the
$180^\circ$-reorientational jumps \cite{ref17}, is presently not
clear.

Acknowledgements: We thank the DFG, through SFB 262, for financial
support. Part of this work was done on the computer facilities of the
Regionales Rechenzentrum Kaisers\-lautern.

\newpage

\begin{figure}[f]
\psfig{file=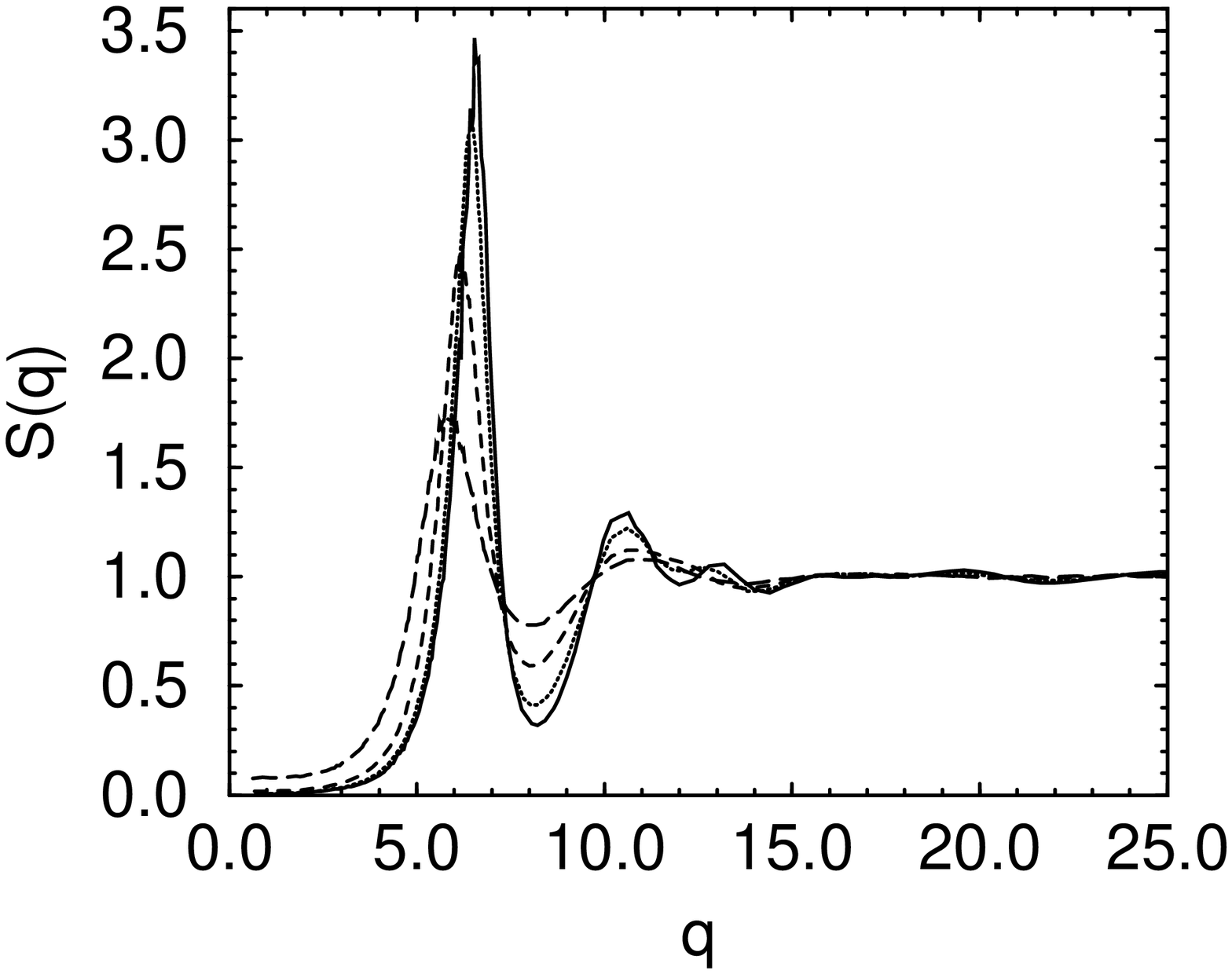,width=13cm,height=9cm}
\caption{
Wave vector dependence of the static structure factor $S(q)$ 
(center of mass) for $T$ = 0.477, 
0.63, 1.1 and 2.0 (at the main peak from top to bottom).}
\label{fig1}
\end{figure}
\begin{figure}[f]
\psfig{file=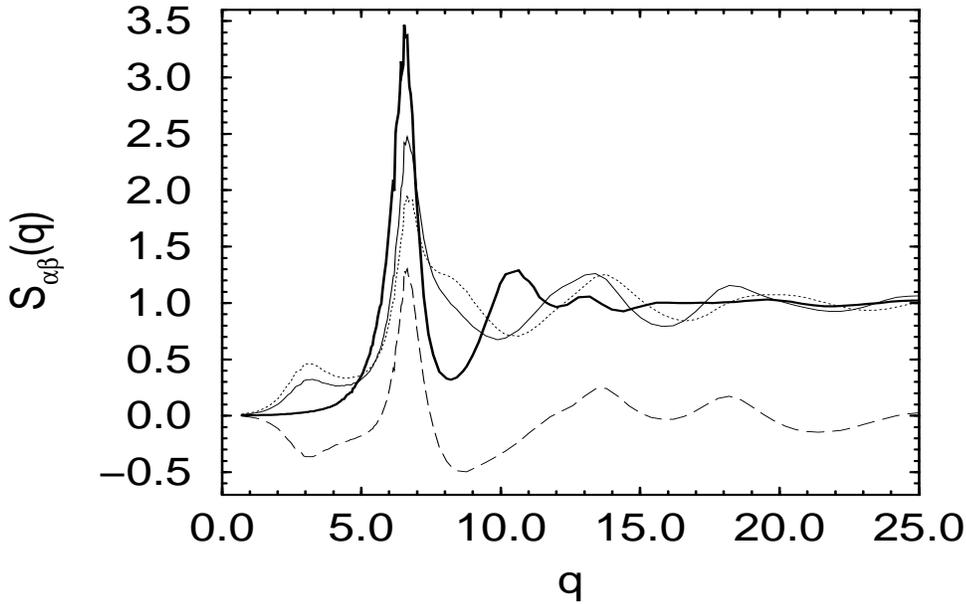,width=13cm,height=9cm}
\caption{
$S(q)$ (bold solid line) and the partial structure factors 
$S_{\alpha \beta}$ 
versus $q$ for $T$ = 0.477; $S_{AA}$ (solid line), $S_{BB}$ (dotted line) 
and $S_{AB}$ (dashed line).}
\label{fig2}
\end{figure}
\begin{figure}[f]
\psfig{file=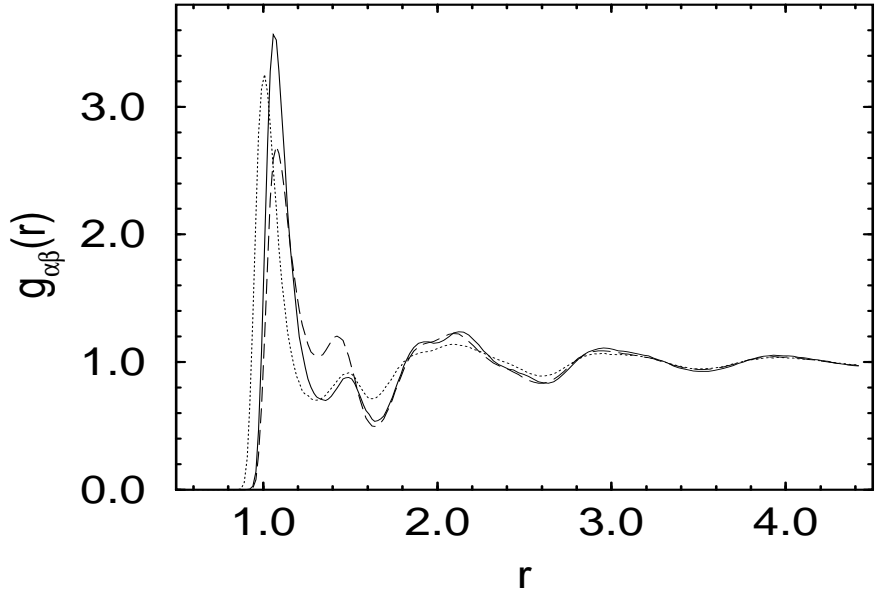,width=13cm,height=9cm}
\caption{
$r$-dependence of the atomic radial pair distribution function
$g_{\alpha \beta}(r)$ for $T$ = 0.477; $g_{AA}$ (solid line), $g_{BB}$
(dotted line) and $g_{AB}$ (dashed line)}
\label{fig3}
\end{figure}
\begin{figure}[f]
\psfig{file=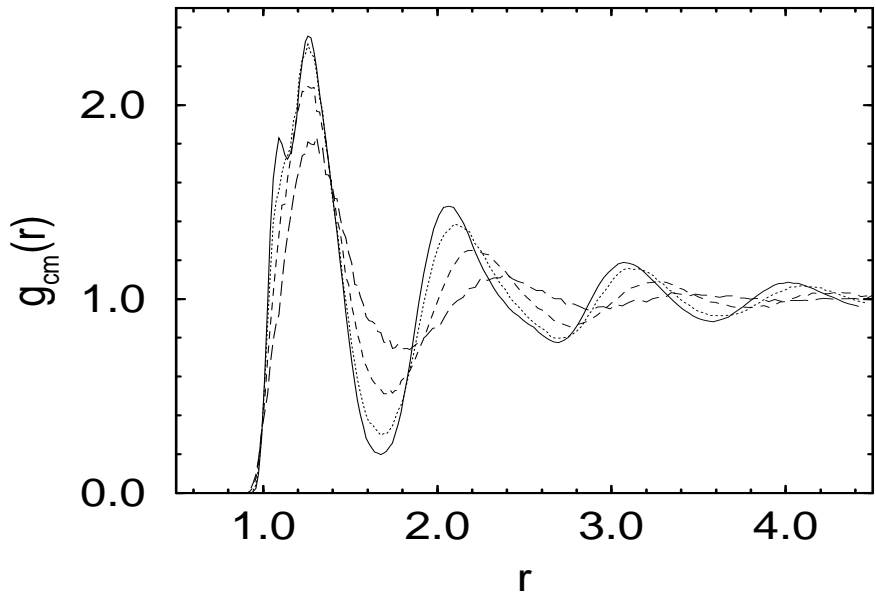,width=13cm,height=9cm}
\caption{
Radial distribution function $g_{cm}(r)$ for the center of 
mass positions and for $T$ = 0.477 (solid line), 0.63 (dotted line), 1.1 
(short dashed line) and 2.0 (long dashed line).}
\label{fig4}
\end{figure}
\begin{figure}[f]
\psfig{file=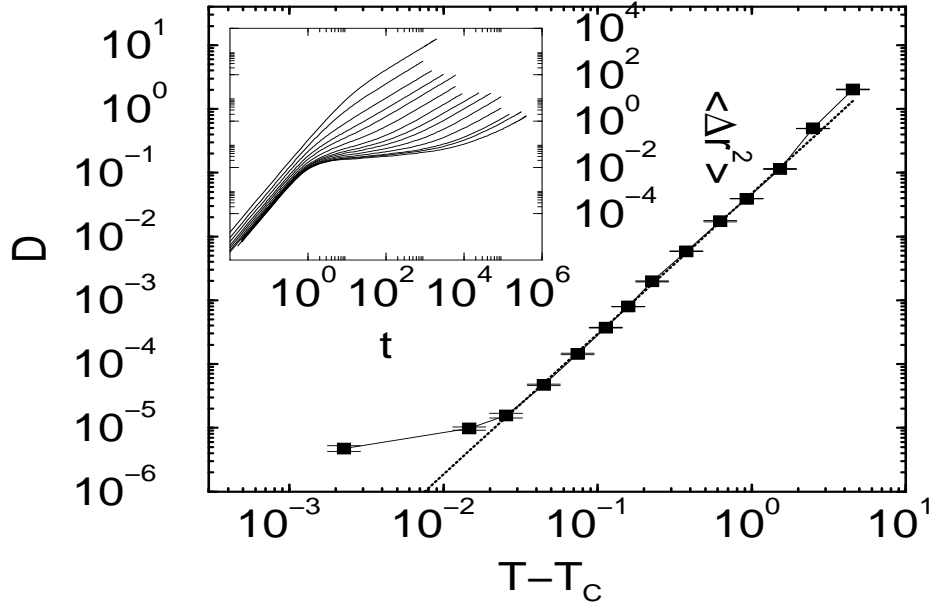,width=13cm,height=9cm}
\caption{
Temperature-dependence of the translational diffusion constant $D$;
numerical data (squares, including error bars), power law (dotted line).
The solid line serves as a guide for the eye. Inset: time
dependence of the mean squared displacement for all 
temperatures investigated.}
\label{fig5}
\end{figure}
\begin{figure}[f]
\psfig{file=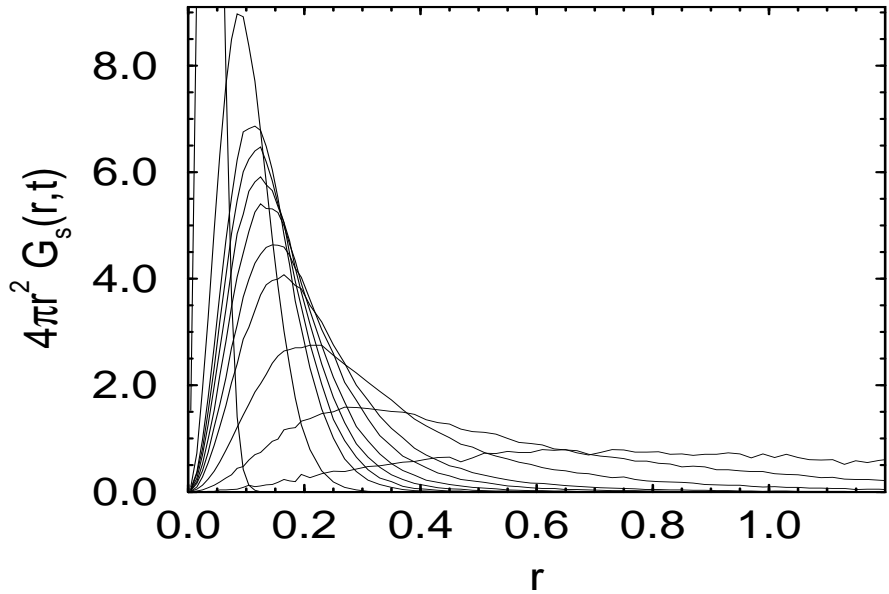,width=13cm,height=9cm}
\caption{
$r$-dependence of $4\pi r^2 G_s(r,t)$ for $t$ = 0.4, 1.7, 5.88, 24.3,
101, 416, 1723, 5000, $2.5\cdot10^4$, $10^5$ and $3.5\cdot 10^5$ (from
top to bottom) and for $T=0.477$.}
\label{fig6}
\end{figure}
\begin{figure}[f]
\psfig{file=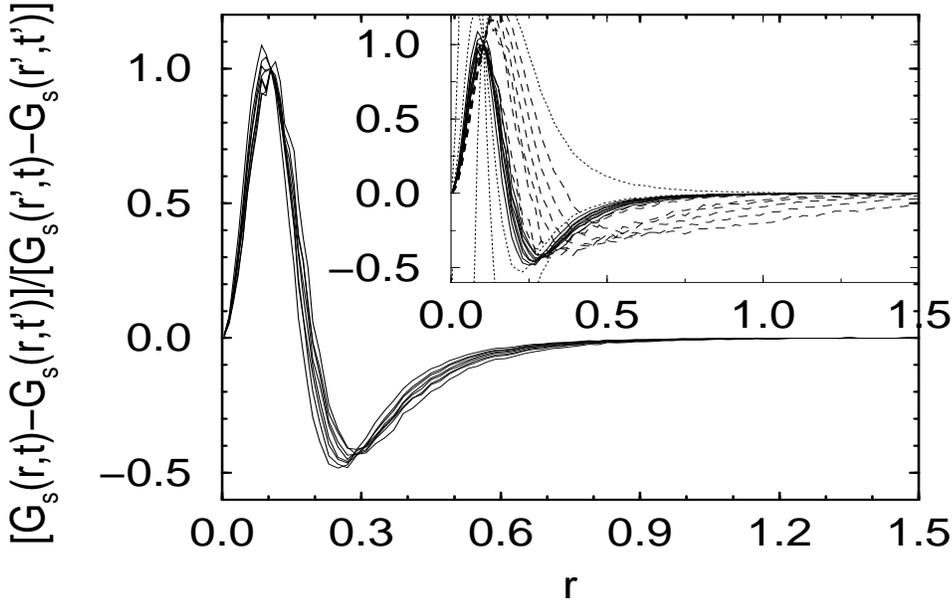,width=13cm,height=9cm}
\caption{
$(G_s(r,t)-G_s(r,t')) / (G_s(r',t)-G_s(r',t'))$ versus $r$ 
for fixed $r'$ = 0.105 and $t'$ = 5000 for $3.44 \leq t \leq 1723$ and for 
$T = 0.477$. Inset: this 
ratio for the same $r'$, $t'$ and $T$, but for a much larger time span 
$0.4 \leq t \leq 3.5 \cdot 10^5$. The times $t$ are equidistant on a 
logarithmic time axis:$t \approx 0.40 \cdot 2^n$ with $n=0,1,2,\ldots$.
Dotted lines: short times; solid lines: $\beta$-relaxation regime;
dashed lines: long times.}
\label{fig7}
\end{figure}
\begin{figure}[f]
\psfig{file=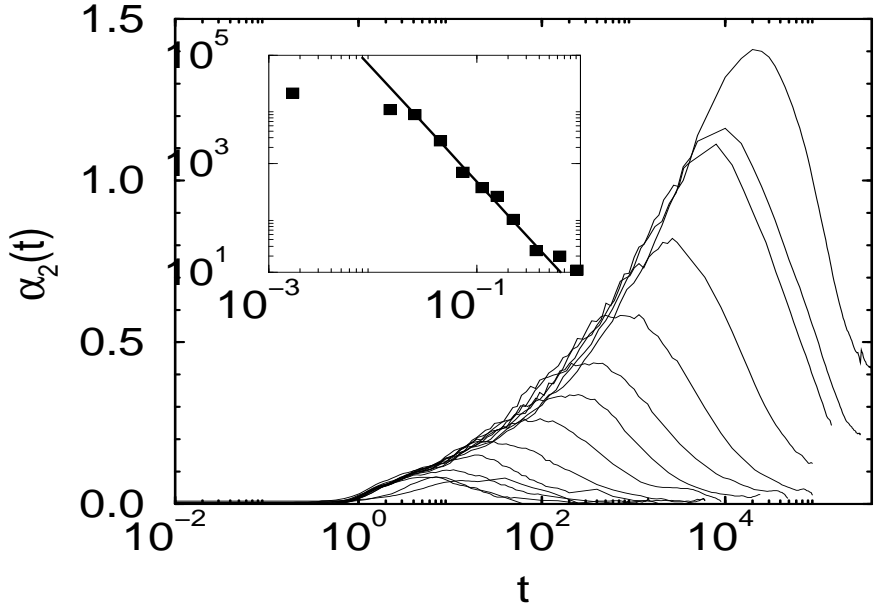,width=13cm,height=9cm}
\caption{
The non-Gaussian parameter $\alpha_2(t)$ versus time for all
temperatures investigated. Inset: the maximum position $t_{max}$
versus $T-T_c$ with $T_c=0.475$.}
\label{fig8}
\end{figure}
\begin{figure}[f]
\psfig{file=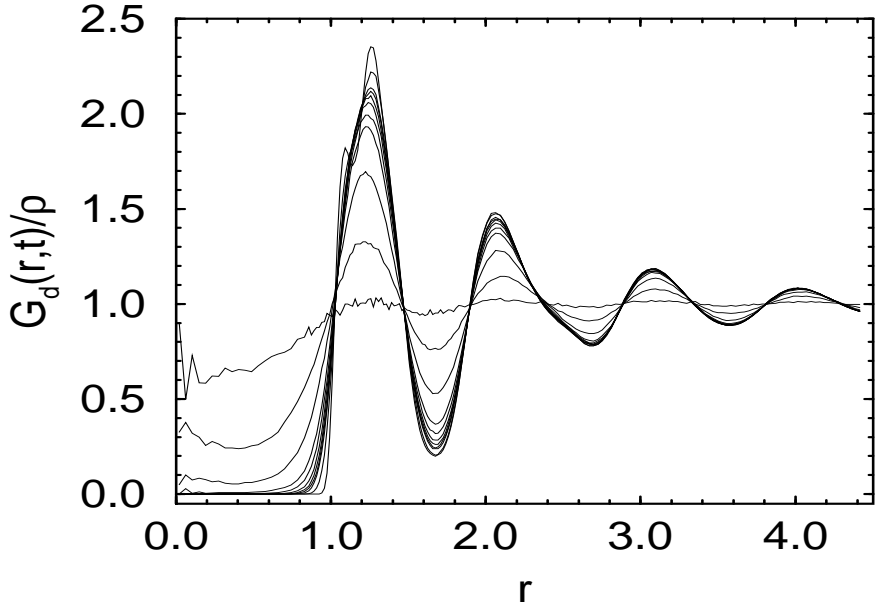,width=13cm,height=9cm}
\caption{
$r$-dependence of $G_d$, normalized by the average density 
$\rho$, for the same times as in Fig. 6 and for $T$ = 0.477.}
\label{fig9}
\end{figure}
\begin{figure}[f]
\psfig{file=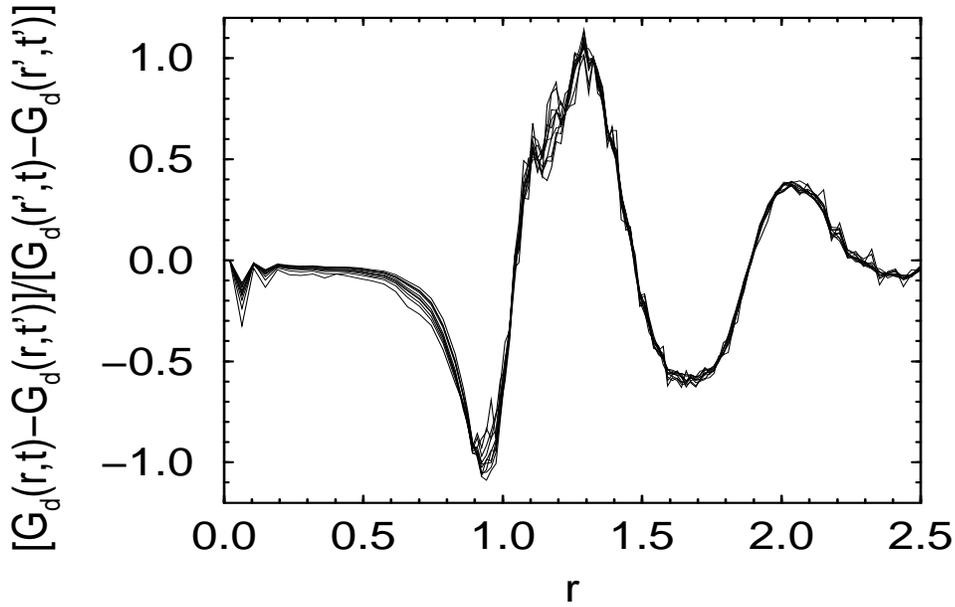,width=13cm,height=9cm}
\caption{
$(G_d(r,t)-G_d(r,t')) / (G_d(r',t)-G_d(r',t'))$ versus $r$ 
for fixed $r'$ = 1.325 and $t'$ = 5000 for the same times $t$ as in Fig.~7 
and for T = 0.477.}
\label{fig10}
\end{figure}
\begin{figure}[f]
\psfig{file=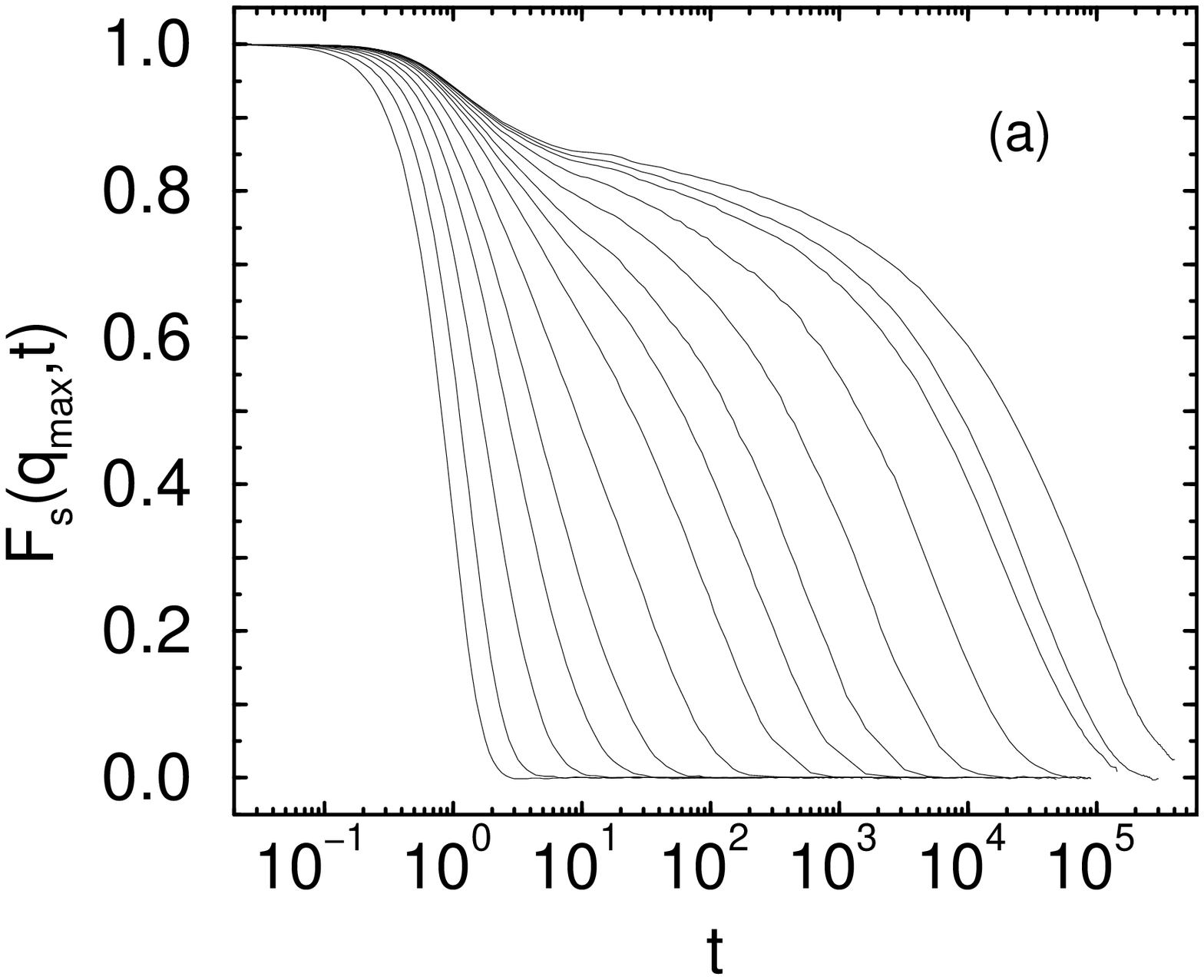,width=13cm,height=9cm}
\end{figure}
\begin{figure}[f]
\psfig{file=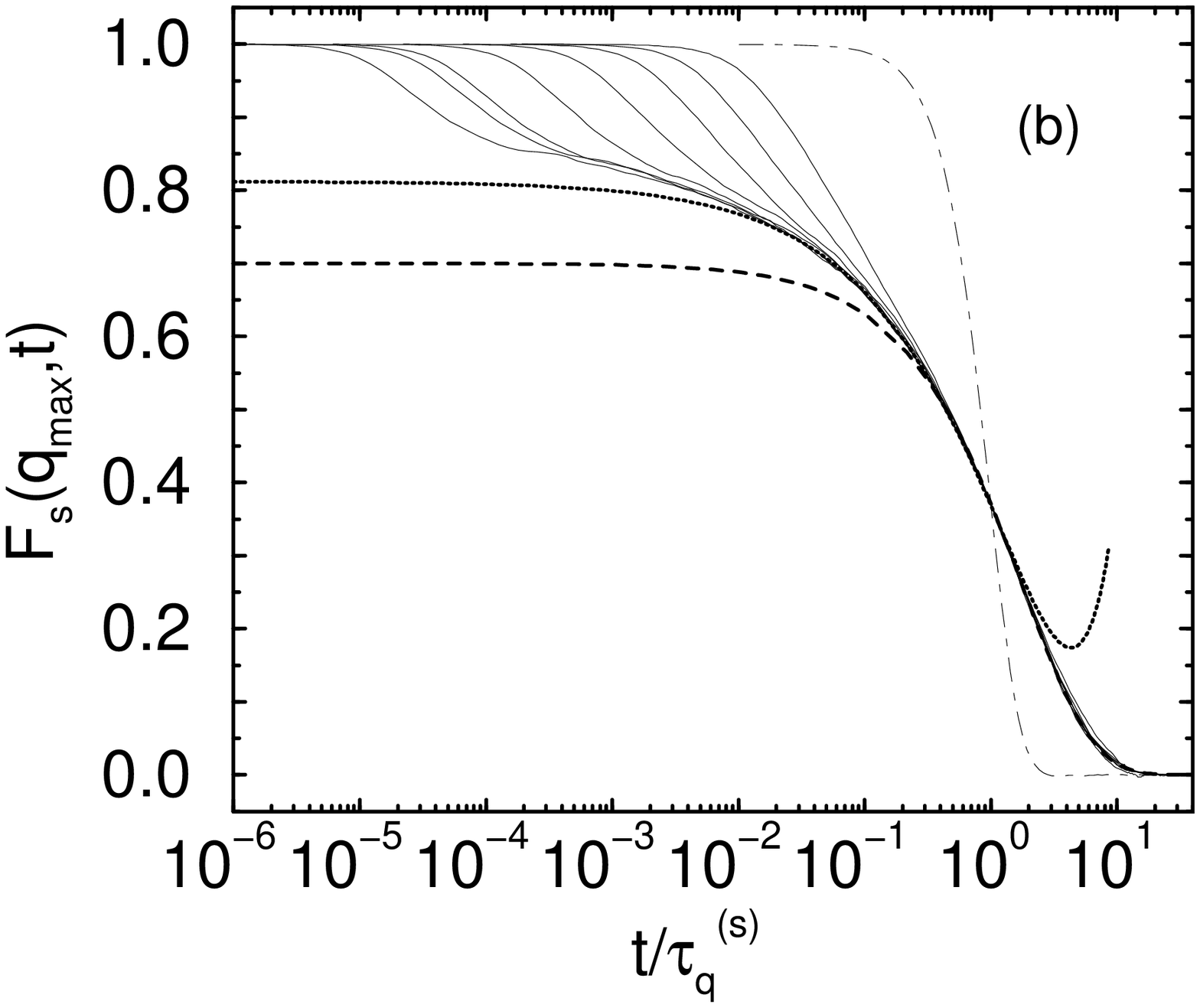,width=13cm,height=9cm}
\caption{
(a) Time-dependence of the self part $F_s$ for $q=q_{max}=6.5$ for all
temperatures investigated. (b) $F_s(q_{max},t)$ versus rescaled time.
The eight curves (solid lines) on the left refer to the eight lowest
investigated temperatures and the one farthest to the right
(dashed-dotted curve) corresponds to the highest temperature $T$ = 5.0.
The von Schweidler law including corrections (cf. Eq.~9) is represented
by the dotted line, and a KWW-fit is shown dashed.}
\label{fig11}
\end{figure}
\begin{figure}[f]
\psfig{file=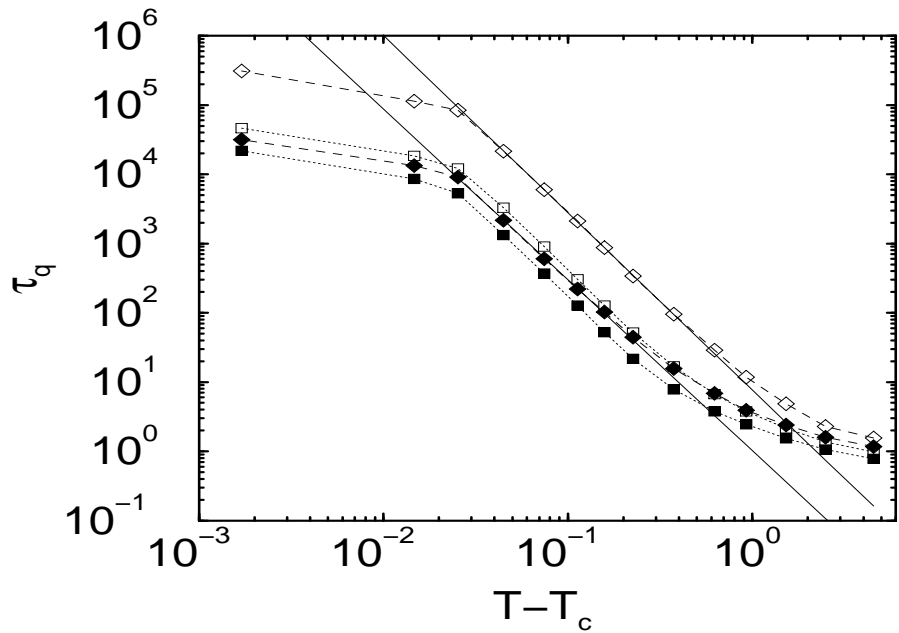,width=13cm,height=9cm}
\caption{
$\alpha$-relaxation times $\tau_q$ versus $T-T_c$: open diamonds from
$F(q_{max},t)$, filled diamonds from $F(q_{min},t)$, open squares from
$F_s(q_{max},t)$, filled squares from $F_s(q_{min},t)$. The dashed
lines are a guide for the eye and both solid lines represent fits with
a power law.}
\label{fig12}
\end{figure}
\begin{figure}[f]
\psfig{file=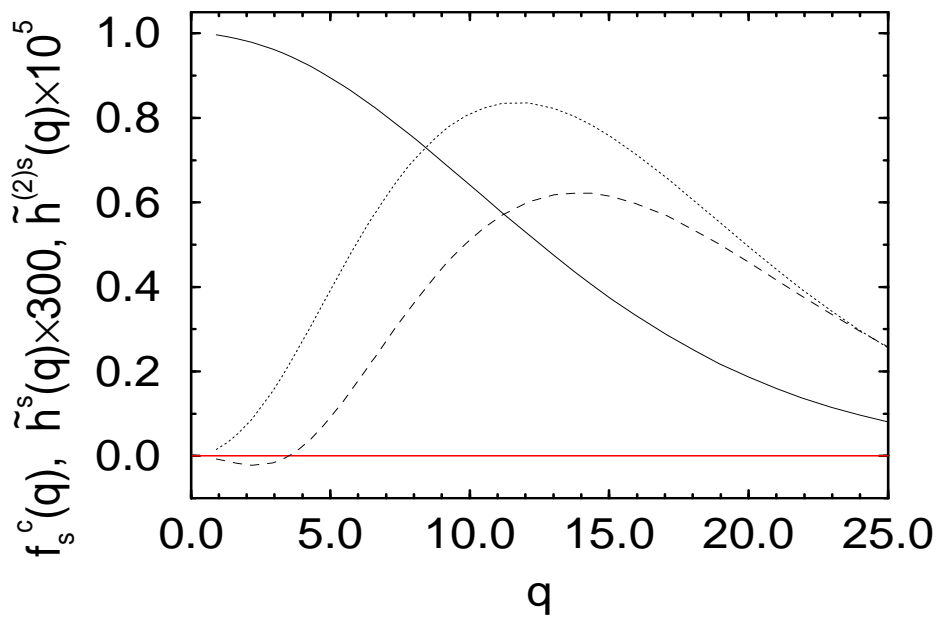,width=13cm,height=9cm}
\caption{
Wave vector dependence of $f_s^c(q)$ (solid line), critical amplitude
$\tilde{h}^{s}(q)$ (dotted line) and the correction $\tilde{h}^{(2)s}(q)$ 
(dashed line) versus $q$.}
\label{fig13}
\end{figure}
\begin{figure}[f]
\psfig{file=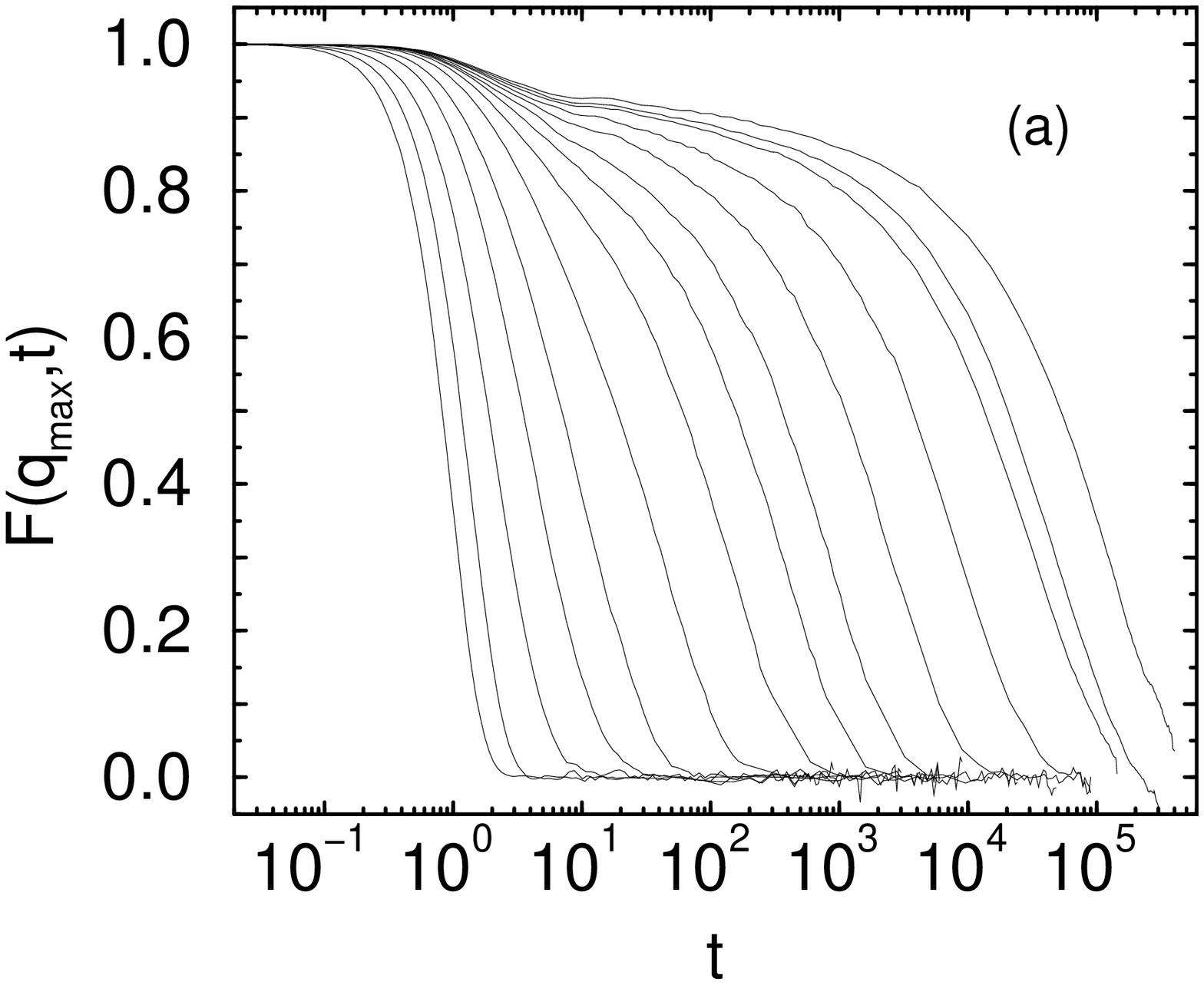,width=13cm,height=9cm}
\end{figure}

\begin{figure}[f]
\psfig{file=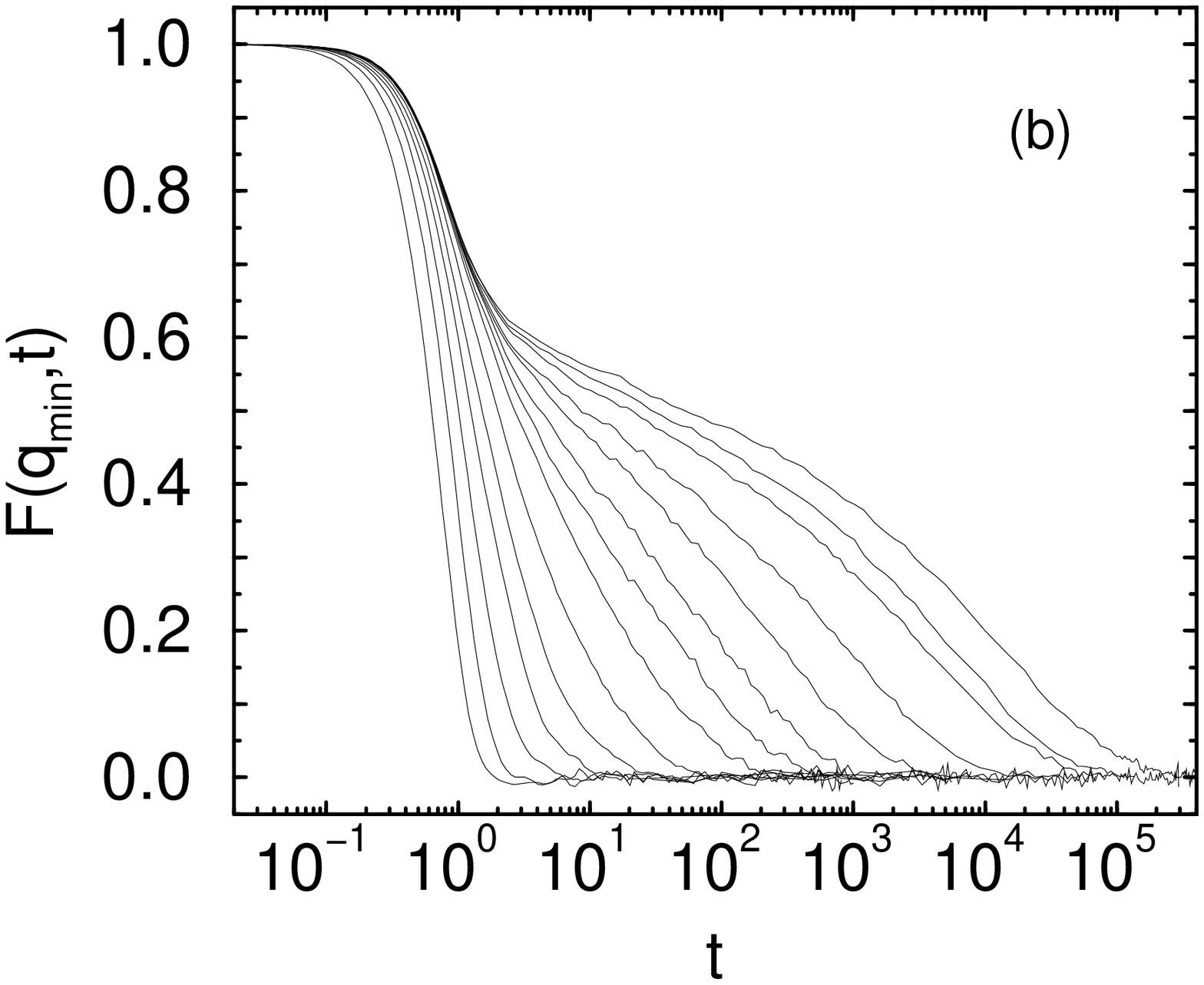,width=13cm,height=9cm}
\caption{
Time-dependence of $F(q,t)$ for all temperatures investigated. (a)
$q=q_{max}=6.5$, (b) $q=q_{min}=8.15$.}
\label{fig15}
\end{figure}
\begin{figure}[f]
\psfig{file=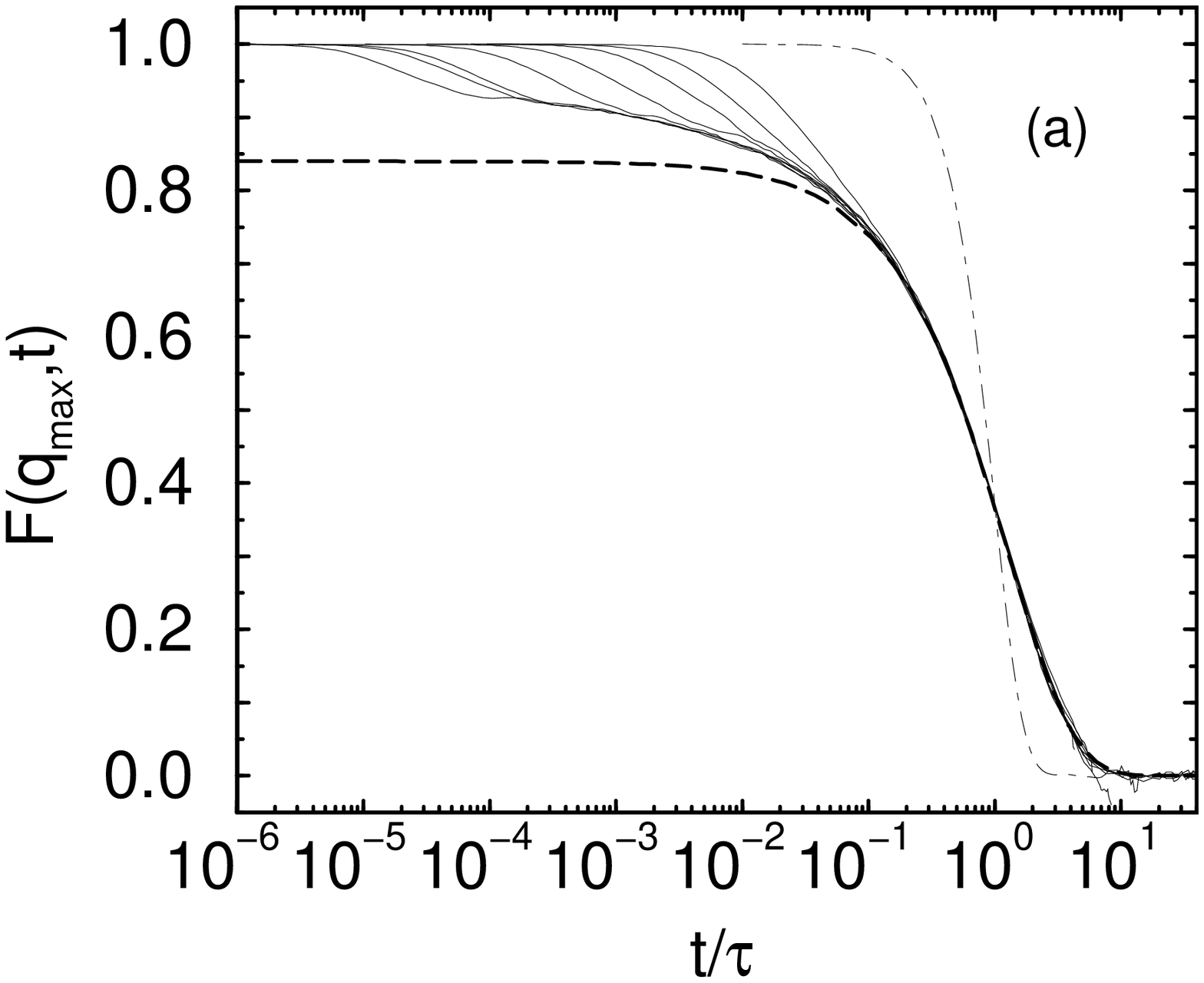,width=13cm,height=9cm}
\end{figure}
\begin{figure}[f]
\psfig{file=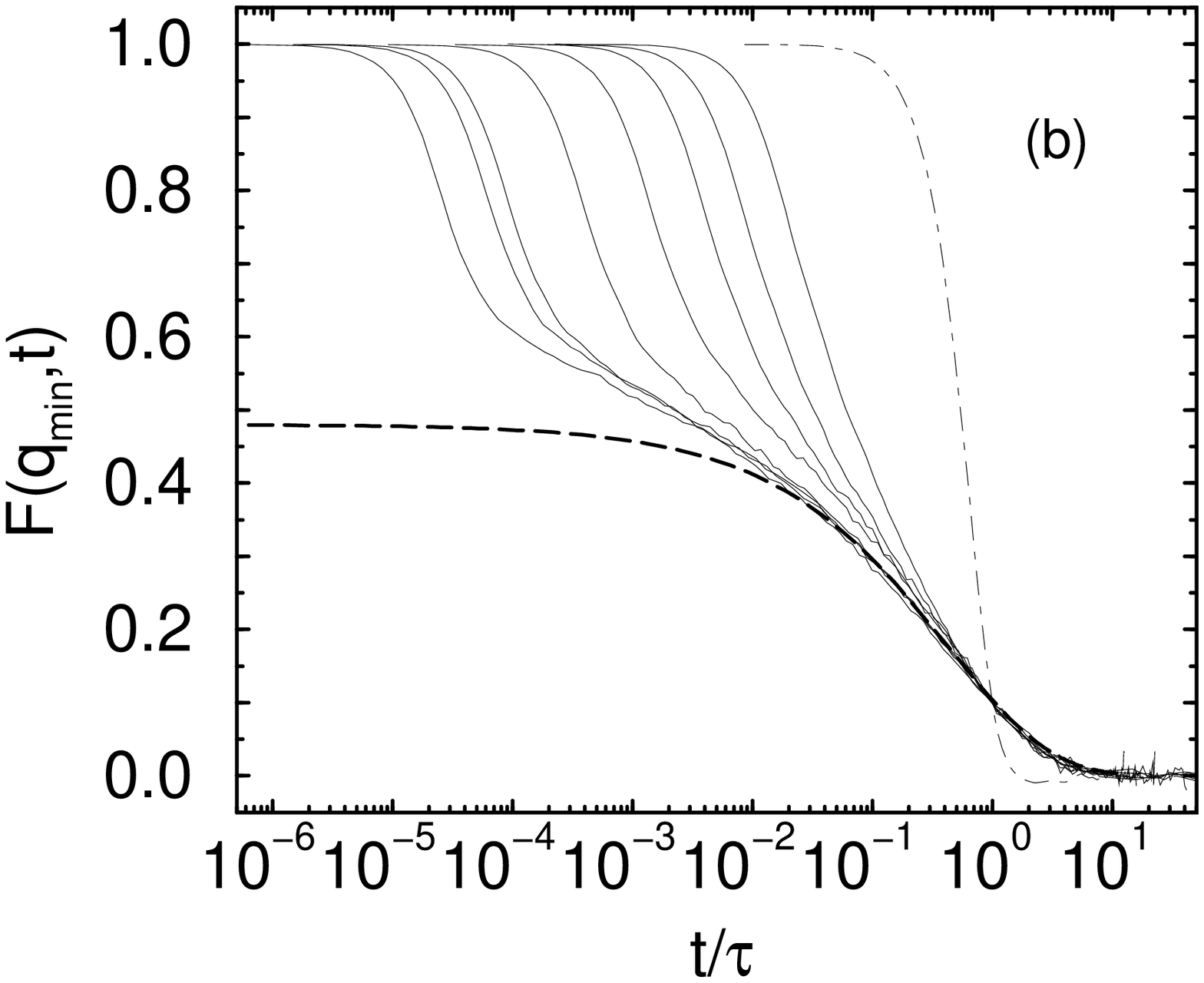,width=13cm,height=9cm}
\caption{
$F(q_{max},t)$ (a) and  $F(q_{min},t)$ (b) versus rescaled time.  The
eight curves (solid lines) on the left refer to the eight lowest
temperatures and the one farthest to the right (dashed-dotted curve)
corresponds to the highest temperature $T$ = 5.0.  The dashed line is a
fit with a KWW-law.}
\label{fig16}
\end{figure}
\begin{figure}[f]
\psfig{file=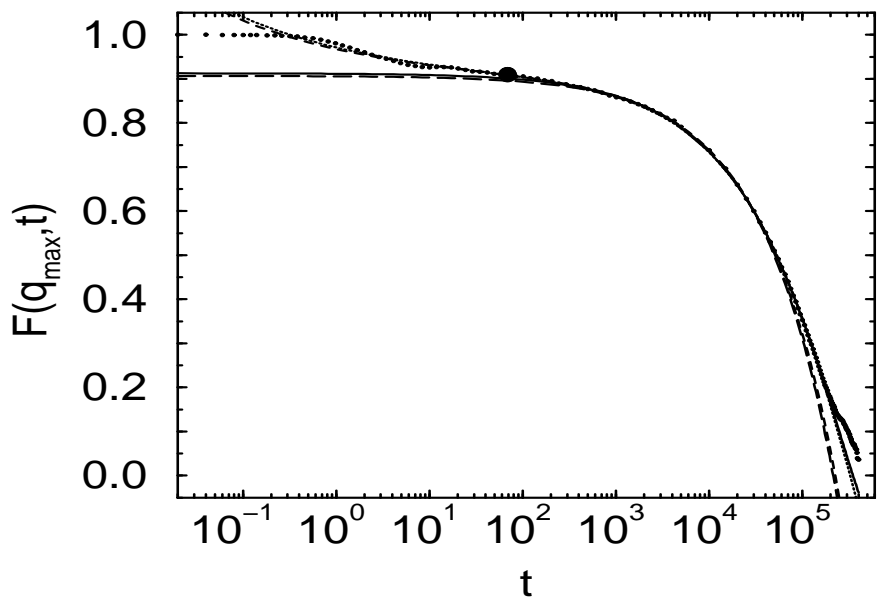,width=13cm,height=9cm}
\caption{
$F(q_{max},t)$ versus $t$ for $T=0.477$: numerical data (bold dots),
$\beta$-correlator with $\lambda$=0.76 (short dashed line),
$\beta$-correlator with $\lambda$= 0.76 including the next order
corrections in the von Schweidler regime (dotted line), von Schweidler
law with $b$ = 0.55 ($\hat{=} \lambda = 0.755$) (long dashed line), 
von Schweidler law with $b$ = 0.55 including the next order 
corrections (solid line).}
\label{fig17}
\end{figure}
\begin{figure}[f]
\psfig{file=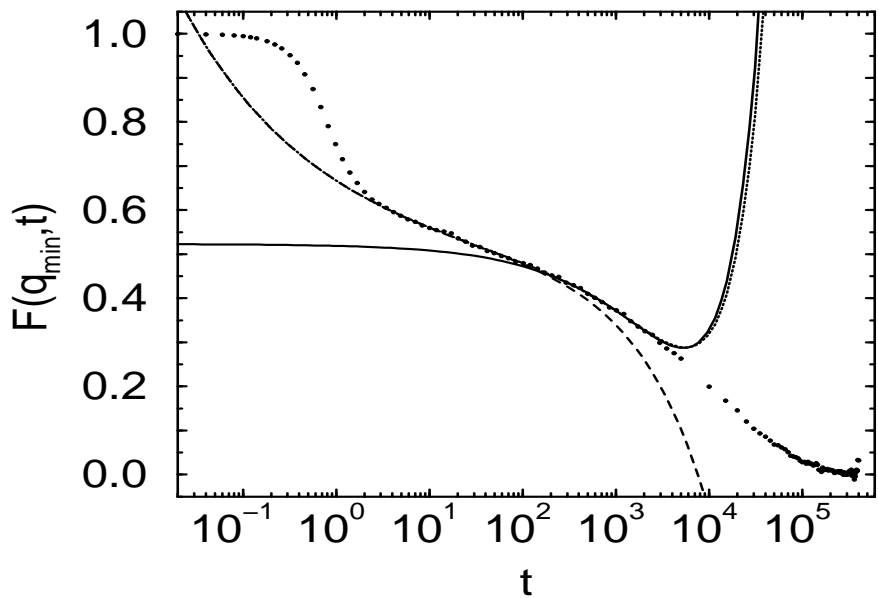,width=13cm,height=9cm}
\caption{
$F(q_{min},t)$ versus $t$ for $T=0.477$: numerical data (bold dots),
$\beta$-correlator with $\lambda$=0.76 (short dashed line),
$\beta$-correlator with $\lambda$= 0.76 including the next order
corrections in the von Schweidler regime (dotted line), von Schweidler
law with $b$ = 0.55 ($\hat{=} \lambda = 0.755$) including the next 
order corrections (solid line).}
\label{fig18}
\end{figure}
\begin{figure}[f]
\psfig{file=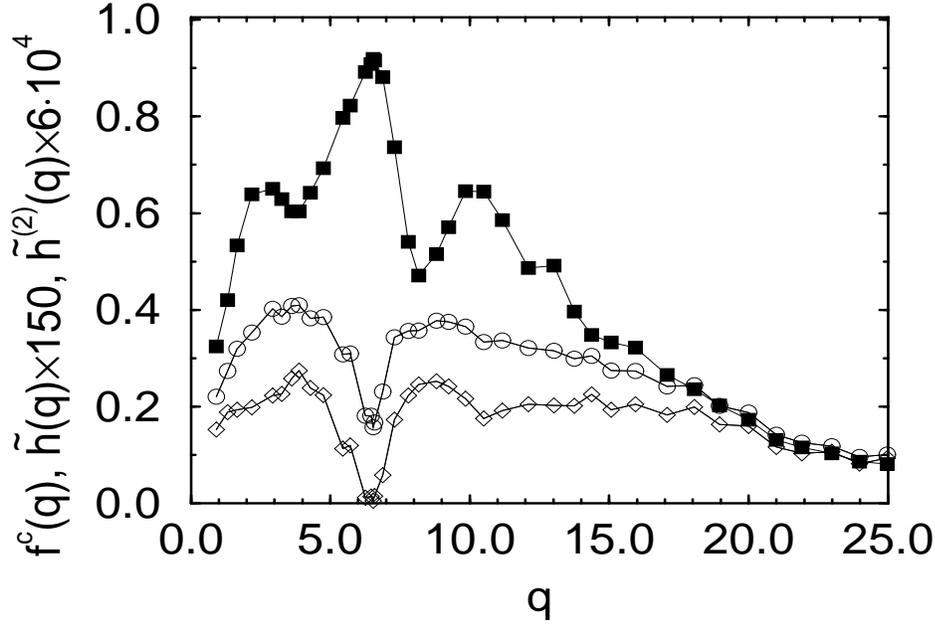,width=13cm,height=9cm}
\caption{
Wave vector dependence of $f^c(q)$ (filled squares), $\tilde{h}(q)$ (open
circles) and $\tilde{h}^{(2)}(q)$ (open diamonds).}
\label{fig19}
\end{figure}
\begin{figure}[f]
\psfig{file=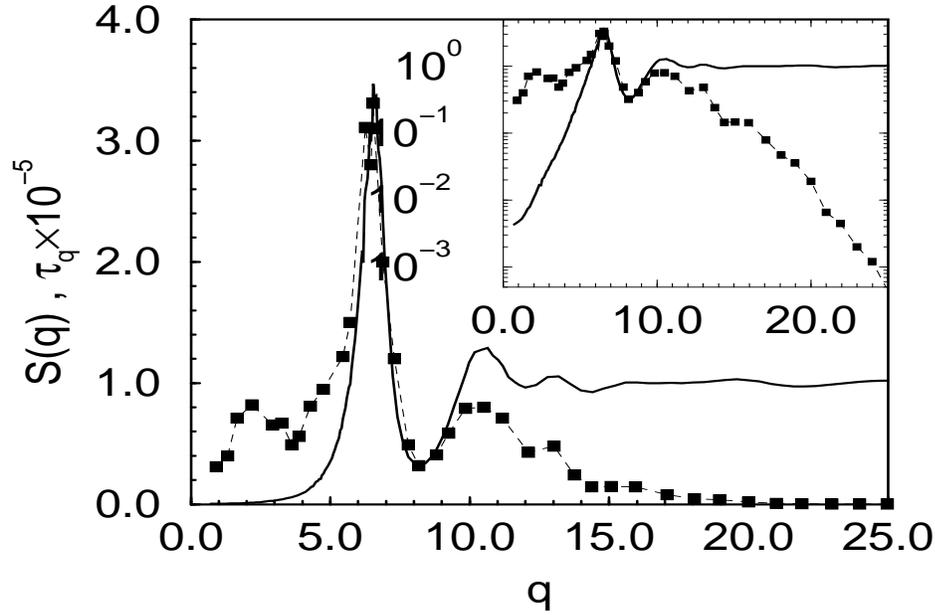,width=13cm,height=9cm}
\caption{
Comparison of $S(q)$ (solid line) for $T$ = 0.477 and the
$\alpha$-relaxation time $\tau_q$ (filled squares) for $T$ = 0.477,
where $\tau_q$ has been multiplied by $10^{-5}$; inset: $\log S(q)$ and
$\log \tau_q$ versus $q$.}
\label{fig20}
\end{figure}
\end{document}